\documentclass[aps,prl,showpacs,twocolumn,groupedaddress,amssymb]{revtex4}

\usepackage{graphicx}
\usepackage{dcolumn}
\usepackage{bm}

\begin{document}


\title{Entropy Maximization and Instability in Uniformly Magnetized Plasma}

\author{S. Son}
\affiliation{18 Caleb Lane, Princeton, NJ 08540}
\author{Sung Joon Moon}
\affiliation{28 Benjamin Rush Lane, Princeton, NJ 08540}

\date{\today}

\begin{abstract}
A regime where a uniformly magnetized plasma could be unstable to a spatial
perturbation in the magnetic field is explored.
In this regime, a uniformly magnetized state does not maximize the entropy.
The physical implication is discussed in the context of the current generation,
the magnetic reconnection, and the dynamo effect. 
\end{abstract}

\maketitle

A uniformly magnetized plasma is often assumed to be dynamically stable,
which is not vulnerable to a localized spatial perturbation in the magnetic field~\cite{Morrison}.
However, if this is not the case, our premise on the uniform magnetic field
should be re-examined carefully. 

The second law of the thermodynamics~\cite{Second} states that
the equilibrium state of a Hamiltonian system maximizes the system
entropy for the given volume and the total energy. 
The concept of the entropy has been often used in the systems of magnetized
plasmas~\cite{entropy1, entropy2, entropy3}.
The equilibrium configuration of a magnetized one-component electron plasma
can be determined by this principle;
the electron distribution function can be obtained by maximizing the Shannon
entropy~\cite{Shanon1, Shanon2} in the framework of the variational principle,
under the constraints of the system-wide conserved quantities such as
the gross number of the electrons, the volume, the energy, and the magnetic moment.
Consider two sets of plasmas, one under a uniform magnetic field $\mathbf{B}_1 = B_0 \hat{z}$
and the other under a spatially-varying field of $\mathbf{B}_2= B_0(1+\beta \cos(kx)) \hat{z}$,
where $\beta$ is a small number.
If the entropy of the latter system is larger than that of the former,
it would attest that the plasma with a uniform magnetic field may 
transition to a spatially varying state.
The goal of this paper is to demonstrate that this is indeed possible in certain regime.

Let us call the plasma with the uniform (spatially varying) magnetic field by
the plasma \textbf{U} (\textbf{V}), and denote the corresponding entropy by $S_U$ ($S_V$). 
We assume that both plasmas have identical system-wide quantities, including the total energy,
the magnetic moment, the total number of the electrons and the volume.  
The total energy for the plasma \textbf{U} with a uniform magnetic field $B_0 \hat{z} $ is
\begin{equation}
  E_U = \int \frac{B_0^2} { 8 \pi } d^3 \mathbf{x}  + \sum_{i} \frac{1}{2} m_e v_i^2 \mathrm{,} 
\end{equation} 
where $m_e$ is the electron mass and the summation is done over all the electrons.
Here the interaction between the electron magnetic moment and the magnetic field is ignored.
The total magnetic moment is
\begin{equation} 
M_U =  \sum_i  \frac{m_e c}{ B_0} v_i^2 \mathrm{,}
\end{equation}
and the Shannon entropy is
\begin{equation} 
 S_U =  - \sum_k p_q \log (p_q) \mathrm{,}
\end{equation} 
where  $c$ is the speed of light,
$p_q$ is the probability that the plasma is in the state $q$,
and the summation is done over all the possible states. 
Using the variation principle and the Lagrange multipliers $l_a$ and $l_b$,
i.e, $\delta S + l_a \delta E_U + l_b \delta M_U =0 $,
we obtain for each $q$
\begin{equation} 
\delta p_q \left( -\log(p_q) + l_a E_U + l_b M_U \right) = 0 \mathrm{,} 
\end{equation}
which leads to an anisotropic Maxwellian distribution 
\begin{equation}
f_U(\mathbf{v})  = \frac{1}{ \sqrt{2 \pi}^3}\frac{n_0}{v_{U\perp}^2 v_{U\parallel}} 
\exp\left( - \frac{v_x^2 + v_y^2}{2v_{U\perp}^2} - \frac{v_z^2}{ 2 v_{U\parallel}^2} \right)\mathrm{,} 
\end{equation}
where $v_{U\perp} $ ($v_{U\parallel}$) is the perpendicular (parallel; $z$-direction) thermal velocity
and $n_0$ is the electron density.
The temperature in each direction is 
determined by the constraint on  the total energy and the total magnetic moment. 

Let us now consider the second plasma \textbf{V}, with the magnetic field 
of $B_0(1+ \beta \cos(kx)) \hat{z} $ where $\beta \ll 1$. 
Following the same steps, the electron distribution is obtained to be
\begin{equation}
f_V(\mathbf{v},x) =  \frac{1}{ \sqrt{2 \pi}^3}\frac{n_V}{v_{V\perp}(x)^2 v_{V\parallel}} 
\exp\left( - \frac{v_x^2 + v_y^2}{2v_{V\perp}(x)^2} - \frac{v_z^2}{ 2 v_{V\parallel}^2}\right) \mathrm{,}  
\end{equation}
where  $v_{V\perp} $ ($v_{V\parallel}$) is the perpendicular (parallel) thermal velocity
\begin{equation}
 v_{V\perp}(x)^2 = \frac{v_{V\perp}^2}{1+ \frac{b \beta \cos(kx)}{1 + \beta \cos(kx)}} \mathrm{,} 
\end{equation}
where  $ b = \gamma -1 $ with $\gamma = v_{U\perp}^2 /v_{U\parallel}^2$.
Three unknowns, $v_{V\perp} $, $v_{V\parallel}$ and $n_V$, can be determined 
as a function of  $v_{U\perp} $, $v_{U\parallel}$, $n_0$ and $\beta $,
given the constraints on the total number of electrons, the energy and the magnetic moment. 
The invariance in the number of electrons
$\int f_V d^3 \mathbf{v} =  \int f_U d^3 \mathbf{v} $
leads to
\begin{equation} 
  n_V = \frac{n_0}{ 1 + b(b+1)\bar{\beta}^2}\mathrm{,} \label{eq:n}
\end{equation} 
where $ \bar{\beta}^2 = \langle \beta^2 \cos(kx)^2 \rangle = \beta^2/2$, and
the constraint on the total magnetic moment
\begin{equation} 
\int f_V \frac{v_x^2 + v_y^2}{B_0(1+\beta \cos(kx))} = 
\int f_U \frac{v_x^2 + v_y^2}{B_0}
\end{equation}
leads to
\begin{equation}
 v_{V\perp}^2 = \frac{ 1 + b(b+1)\bar{\beta}^2 }{ 1 + (3b^2 + 4b + 1)\bar{\beta}^2} v_{U\perp}^2 \mathrm{.}  \label{eq:vp}
\end{equation}
Lastly, the constraint on the energy is given as 
\begin{equation} 
\frac{B_0^2(1+\bar{\beta}^2)}{8\pi} +\int f_V \frac{1}{2} m_e v^2  
=   \frac{B_0^2}{8\pi} + \int f_V \frac{1}{2} m_e v^2  \mathrm{.}
\end{equation}
This leads to the relationship
\begin{equation} 
v_{V\parallel}^2 = v_{U\parallel}^2\left[1 + \frac{2 v_{U\perp}^2 }{ v_{U\parallel}^2 } (2b +1) \bar{\beta}^2 
- \lambda \bar{\beta}^2\right]\mathrm{,}\label{eq:vz}
\end{equation} 
where $\lambda = (B_0^2/8\pi) / (m_e v_{U\parallel}^2/2) $ is the ratio between the 
magnetic energy density and the parallel kinetic energy density of the electron.

Note  that, for a Maxwellian plasma, the Shannon entropy is proportional to 
\begin{equation}
 S \cong  - \log(n_e/v_p^2 v_z) = -\log(n_e) + \log(v_p^2) + \log(v_z) \mathrm{,} 
\end{equation}
where $v_p^2 = v_x^2 + v_y^2$.
Defining   $\delta \log(n_e) = \log(n_V) - \log(n_0)$, $\delta \log(v_p^2) =
 \log(v_{V\perp}^2) -   \log(v_{U\perp}^2) $ and  $\delta \log(v_z) =
 \log(v_{V\parallel}^2) -   \log(v_{U\parallel}^2) $,
we obtain from Eqs.~(\ref{eq:n}), (\ref{eq:vp}) and (\ref{eq:vz}):
\begin{eqnarray} 
 -\delta \log(n_e) &=& b(b+1) \bar{\beta}^2 \mathrm{,} \nonumber \\ \nonumber \\
\delta \log(v_p^2) &=& - (2b^2 + 3b +1) \bar{\beta}^2 \mathrm{,}\label{eq:diff} \\
\nonumber \\ \delta \log(v_z) &=& -\frac{\lambda}{2(1+\lambda)} \bar{\beta}^2
 +  \gamma (2b+1) \bar{\beta}^2 \nonumber \mathrm{.}\\ \nonumber 
\end{eqnarray} 
Finally, the entropy difference between the states $\Delta S \equiv S_V - S_U$ is given by
$\Delta S = 
\left(-(b+1)^2  - \lambda/2 + \gamma (2b+1) \right) \bar{\beta}^2$. 
From  the relationship $\gamma = b+1$, this can be further simplified to  
\begin{equation}
\Delta S = \gamma^2 - \gamma  - \lambda/ 2
\mathrm{,} \label{eq:final2}
 \end{equation}
which is the major result that our following argument is based on.

One important constraint that should be considered is the self-consistency of
the spatially varying magnetic field $B_0 \beta \cos(kx)$  and the current generated by
the magnetic moment $\mathbf{j}_m = \nabla \times \mathbf{m} $, where 
$ \mathbf{m} =\left( \int f_V (mc v_p^2/ B(x)) d^3 v \right)\hat{z} $ is
the magnetic moment density.
Even when the distribution is an isotropic Maxwellian, 
there could be a current in a non-uniform plasma due to the gyro-motion of the electrons. 
In order for the spatially varying magnetic moment to be generated by the current of
the spatially varying magnetic field, 
the relationship  $\delta \mathbf{B} = (4\pi/c) \mathbf{m} $, originated  from
the Maxwell equation $\nabla \times \delta \mathbf{B} = (4\pi/c) \mathbf{j}_m $,
 needs to be imposed.
Equating the relationship to the first order in $\beta$, we obtain 
\begin{equation} 
  \gamma \frac{ 1/2 n_0m_e v_{U\perp}^2 }{ B_0^2 / 8\pi} = 1 \mathrm{,} \label{eq:con}
\end{equation}
which is nothing but $ \lambda =\gamma^2$. 
The condition  $\Delta S >0$  (or $\gamma > 2 $) would be  the condition for the possible  instability.
 In order for our analysis to be valid, the wave vector $k$ should be less than
the inverse of the typical electron gyro-radius so that 
$kv_{U\perp} / \omega_{ce} < 1$, where $ \omega_{ce}$ is the gyro-angular frequency. 
While the unstable regime identified here is the same as for the Weibel instability~\cite{Weibel,Yoon}, 
the physical origin is  different. 
In the Weibel instability, the range of the unstable wave vector is
$0< k <\sqrt{\gamma^2/2 - 1} (\omega_{\mathrm{pe}}/c)$; 
however it is much wider in our theory, $0<k<r_g^{-1}$,
where $r_g$ is the gyro-radius. 
It would be interesting to examine how the possible instability identified 
here is different from the Weibel instability.

Let us consider the constraint $\lambda = \gamma^2$.
It is imposed on the condition that 
the spatially varying part of the magnetic field arises from
the current generated by the electron magnetic moment.
While this is a necessary condition for a self-sustaining system, 
it might be possible a current can be driven from outside,
via various methods~\cite{Fisch,sonprl}, without
perturbing the total energy or the magnetic moment of the plasma
so that the condition can be lifted from the constraint.  
If $\lambda $ is almost zero, then the instability can exist for a nearly
isotropic Maxwellian plasma, as the plasma is unstable even when $\gamma \cong 1$. 

The regime of our interest is where the gyro-frequency is faster than
the time scale at which the magnetic field evolves and the collision frequency is
much slower than the time scale the magnetic field changes.
In this circumstance, the magnetic moment of an individual electron is conserved
under the changing magnetic field, while the energy and the momentum are not;
the magnetic field acts as a storage for the total energy and the total
momentum in the $z$-direction. 
For this reason, we choose the total energy, the total number of electrons and
the total magnetic moment to be the constraint. 
In principle, the momentum in the $z$-direction needs to be considered as a constraint.
With $f_U$ having a drift $v_{z0}$ and $f_V$ having a local drift $v_{z}(x)$,
$v_{z}(x) $ can be determined from the momentum constraint, 
as a function of $v_{U\perp} $, $v_{U\parallel}$, $n_0$, $v_{z0}$ and $\beta $.
However, the drift $v_{z0}(x)$ has no impact on the entropy and the result given above
still holds even if the momentum conservation is imposed as an additional constraint. 
In our work, it is assumed that the total momentum in the $z$-direction is zero
(from $f_U$ and $f_V$) because the non-zero total momentum does not alter the analysis given above. 

It should be noted that as the magnetic moment 
is not conserved in the presence of the collisions,
the result presented here is valid only in the time scale faster than 
the electron collision rate. 
In a collisionless system, the magnetic moment \textit{distribution},
as well as the total magnetic moment itself, is conserved.
As the plasmas \textbf{U} and \textbf{V} do not have the same \textit{distribution}, 
the possibility of the dynamic connectivity between \textbf{U} and \textbf{V} should be considered.
One important point to be noted is that 
due to the spatial variation of the magnetic field,   
an electron's magnetic moment is ambiguous up to the order of  $r_g k$.
As an illustrative example,
consider  $\mathbf{B}(x) = B_{+}\hat{z}$ for $x>0$ and $\mathbf{B}(x) =B_{-}\hat{z}$ for $x<0$.  
An electron with the gyro-center at $x=0$ 
can have different magnetic moment depending on the traveling direction ($x>0$ or  $x<0$).
Although the total magnetic moment is conserved (in the statistical sense),
the final detailed distribution of the magnetic moment is path-dependent
upon how  the magnetic field configuration  changes from 
$B_0 \hat{z}$  to   $B(1+\beta cos(kx)) \hat{z}$. 
Consider two cases where the magnetic field changes from $B_0 \hat{z}$ to 
$B_0(1+\beta_2 cos(2kx)) \hat{z} $ and then finally to $B_0(1+\beta cos(2x)) \hat{z}$,
and where it changes from $B_0 \hat{z}$ directly to   $B_0(1+\beta cos(kx)) \hat{z}$.
The \textit{distributions} of these two cases cannot be the same.
In other words, 
there are an infinite degrees of freedom for the paths among which
there might be one eventually connecting the plasma \textbf{U} and the plasma \textbf{V}.
While the above argument does not guarantee 
the dynamic connectivity between \textbf{U} and \textbf{V},
it shows that the chance that there exists a path from the plasma \textbf{U} to the plasma
\textbf{V$^\prime$}, where the plasma \textbf{V$^\prime$} is very close to the plasma \textbf{V},
is high.
The question should be addressed in more detail, in the framework of the gyro-kinetic treatment~\cite{hahm, littlejohn}. 
It is interesting and fundamentally important question in the context given in our work,
which is beyond the scope of this paper. 
 
With the caution on the dynamical connectivity discussed above, 
let us assume that the plasmas \textbf{U} and \textbf{V} are dynamically connected.
Then our theory predicts that the uniform Maxwellian plasma can be unstable in certain regime.   
As the plasma gets squeezed or expanded as in the inertial confinement fusion process~\cite{tabak},
the z-pinch plasma or solar corona magnetic reconnection region~\cite{corona}, 
the plasma may cross the boundary given in Eq.~(\ref{eq:con}) with $\gamma > 2 $.
If it does occur, as it crosses the boundary, the plasma would undergo one of the following
three transitions before reaching to a stable state.
First, the plasma may develop a spatially-varying magnetic field, as considered in our work. 
This is possible only with the spatially-varying current generation. 
Second, the plasma may convert the magnetic field energy into the electron kinetic energy,
as in the case of the magnetic reconnection. 
Finally, the plasma may convert the electron kinetic energy into the magnetic field energy,
as in the case of the dynamo effect. 
Our analysis does not provide the answer which transition should occur.

When the magnetic field is highly intense, as is often the case in the astrophysical plasma,
the quantum Landau level and other quantum effects become significant~\cite{sonpla, sonprl, sonlandau}.
In such a case, the ground state may not be at its minimum energy or the maximum entropy level
when the magnetic field is uniform. This could lead to interesting phenomena.
Complication would be the quantum diffraction and degeneracy~\cite{sonprl, sonbackward, sonlandau, sonpla}. 

We would like to thank Prof. Fisch on various helpful discussion with him on the subject.

\end{document}